\documentclass[a4paper]{jpconf}
\usepackage{graphicx}
\begin{document}
\title{Inclusive Open Charm Production in pp
and Pb--Pb collisions with the ALICE Detector}

\author{Renu Bala, for the ALICE Collaboration}

\address{INFN, Torino, Italy}

\ead{bala@to.infn.it}

\begin{abstract}
ALICE is the dedicated heavy-ion experiment at the LHC. Its main
physics goal is to study the properties of  strongly-interacting matter at conditions of high energy density ($>$10 GeV/$\rm fm^3$) and high
temperature ($>$ 0.5 GeV) expected to be reached in central Pb–Pb
collisions. Charm and beauty quarks are  powerful tools to investigate this high density and strongly interacting state of matter since they
are produced in initial hard scatterings that are therefore generated
early in the system evolution and probe its hottest, densest stage.
 The measurement
of the charm production cross sections in pp collisions provides an interesting
insight into QCD processes and is crucial as a reference for heavy
ion studies. We present  open charm cross section
measurements in pp collisions at $\rm \sqrt{s}$ = 7 TeV and
$\rm \sqrt{s}$=2.76 TeV in the central rapidity region.
In addition, the first measurement of nuclear modification factor of D-meson in
Pb--Pb collisions at $\rm \sqrt{s}$= 2.76 TeV is shown.
\end{abstract}

\section{Introduction}

Heavy quarks are unique probes to
study the Quark-Gluon Plasma produced in heavy ion collisions at the LHC.  Due to their large masses, they are produced
predominantly  in hard scatterings,  during the
initial phase of the collision. Therefore, they can probe the
properties of the strongly interacting matter during its hottest,
densest phase. One of the key methods used to infer the parameters of the medium is the measurement of energy loss of the partons traversing
it. In a QCD picture, radiative in-medium energy loss is one of the
main mechanisms expected to contribute, with dependence
 on the mass and the colour charge of the particle. The
radiation is suppressed at small angles for massive partons because of
the dead-cone effect
\cite{dead-cone} and is larger for gluons, which have stronger color
charge with respect to quarks (Casimir effect). Therefore
one should observe a pattern of decreasing energy loss when going from
the light flavour hadrons ($\rm h^{\pm}$ or
$\pi^{0}$) which mainly originate in gluon jets to  D and  B-mesons. The measurement and comparison
of these different  probes of the medium provides a unique test of the colour
charge and mass dependence of parton energy loss \cite{energy loss}.

Heavy-quark production measurements in pp collisions at the LHC, besides providing a necessary
reference for the study of medium effects in Pb--Pb collisions, are interesting per se, as a
test of perturbative QCD (pQCD) in a new energy domain, up to 3.5 times above the present energy frontier
at the Tevatron. The charm production cross section measured in
$\rm p\bar{p}$ collisions at $\rm \sqrt{s}$=1.96 TeV at the Tevatron \cite{CDF}
is at the upper limit of the uncertainty band in current pQCD
caculations, as 
observed also in pp collisions at RHIC at the much lower energy of $\rm \sqrt{s}$=
0.2 TeV \cite{STAR}.

\section{ALICE Detector}

The ALICE detector \cite{JINST}  consists of two parts:  a central
barrel at mid-rapidity  and a muon spectrometer at forward
rapidity. For the  present analysis,  we have used the information from a
subset of the central barrel detector, namely the Inner Tracking System
(ITS), the Time Projection Chamber (TPC), the Time Of Flight
detector (TOF), the T0 detector for time zero measurement  and the VZERO scintillator
arrays for triggering. The two tracking
detectors, the ITS  and the TPC, enable the reconstruction of 
charged particle tracks in the pseudorapidity range -0.9 $ \rm < \eta
<$0.9 with a  momentum resolution better than 4$\%$ for $\rm p_{t} <
$20 GeV/c and provide charged particle identification via a dE/dx measurements.
The ITS  is a crucial detector for open heavy flavour studies because it
allows the track  impact parameter (i.e. the distance of
closest approach of the track to the primary vertex) to be measured  with a resolution
better than 75 $\rm \mu$m for $p_{t} > $ 1 GeV/c,  thus providing
the capability to detect the secondary vertices originating from
heavy-flavour decays. The TOF detector provides  particle identification by
time of flight measurement.

The results that we present are  obtained from data recorded with a
minimum bias trigger  in pp and Pb--Pb collisions. The trigger was defined by the
presence of a signal at least one  of two scintillator arrays,
positioned in the forward and backward regions of the experiment, or in the silicon
pixel barrel detector. For pp collisions, the D-meson  production cross sections are normalized relative to
the minimum-bias trigger cross section, which was measured using a van-der-Meer scan
technique \cite{Ken}. Pb--Pb collision-centrality classes are defined using a Glauber-model analysis
of the sum of amplitudes in the VZERO scintillator detector. The total
number of events analyzed  is  $\rm 100\times10^{6}$  (30$\%$ of 2010
data) for pp at 7 TeV,  $\rm 67\times10^{6}$ (full data) for pp at
2.76 TeV and
$\rm 17\times10^{6}$ for Pb--Pb  at 2.76 TeV.

\section{ D meson  Analysis in pp collisions}

The detection strategy for D mesons at central rapidity is based on
the selection of displaced-vertex topologies, i.e. discrimination  of
tracks from the secondary vertex from those originating in the primary vertex, large
decay length (normalized to its estimated uncertainty), and good alignment between
the reconstructed D meson momentum and 
flight-line. The identification of the charged
kaon and pion in the TPC and TOF detectors helps to further reduce the background at low
$\rm p_{t}$. An invariant-mass analysis is then used to extract the
raw signal yield, which is  then
corrected for detector acceptance and reconstruction efficiency,
which are evaluated using a detailed  detector simulation. A
significant fraction of D mesons (10-15$\%$) comes from
B-meson decays,  and since the tracks coming from secondary D mesons are more displaced from
the primary vertex, due to the relatively long lifetime of B mesons
($\rm c\tau \sim$  460-490
$\rm \mu m$), the selection further enhances their contribution to the raw
yield. To estimate this contribution, we used the FONLL \cite{cacciari} cross section for prompt and secondary
D mesons at $\rm \sqrt{s} $ =7 TeV, since the available statistics were 
not sufficient to use data driven methods, as carried out by the CDF \cite{CDF} collaboration.  This
contribution is  subtracted from the
measured raw $\rm p_{t}$ spectrum, before applying the efficiency correction.
                               
\begin{figure}[h]
\includegraphics[width=12pc,height=14pc]{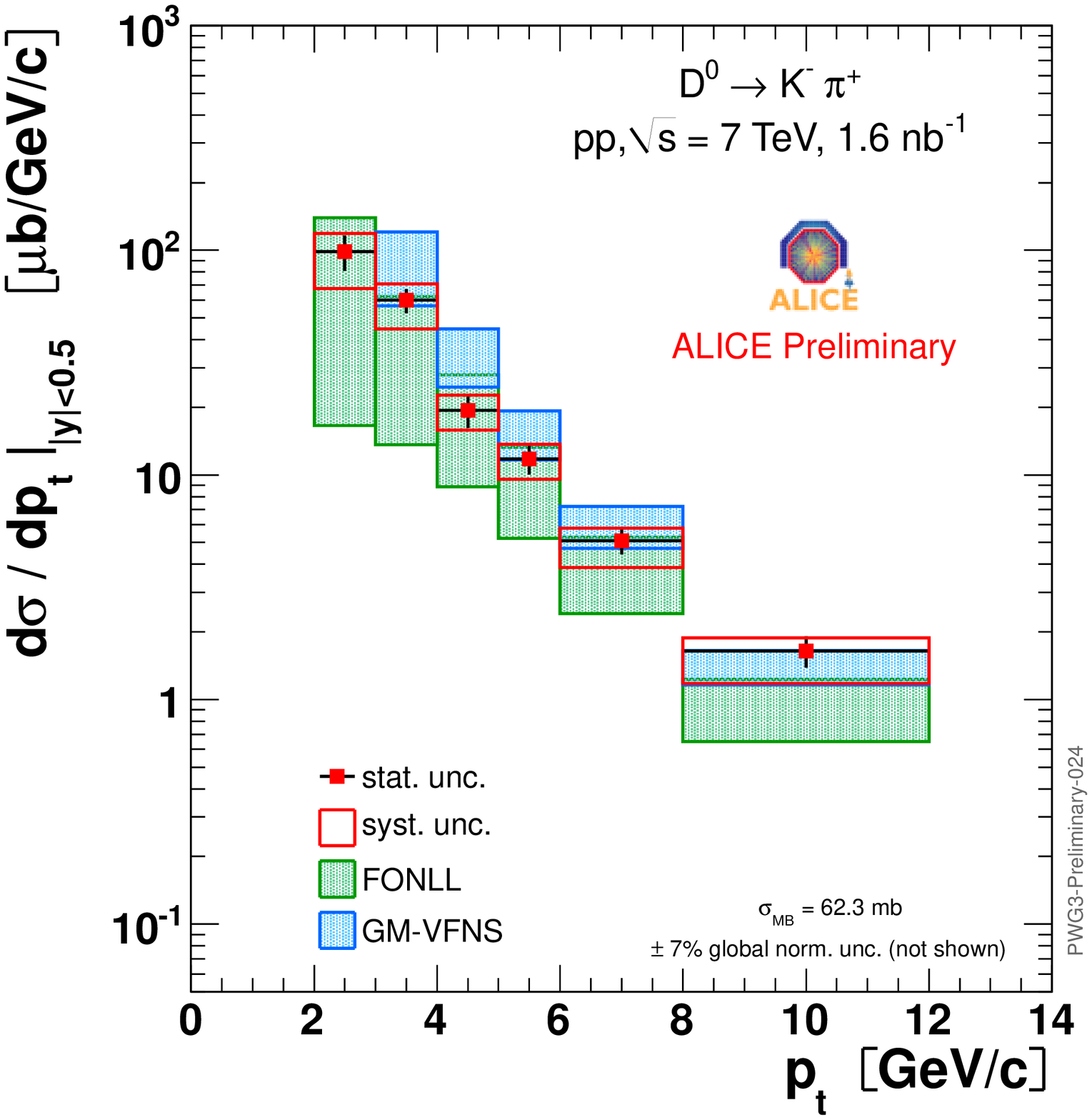}
\includegraphics[width=13pc,height=14pc]{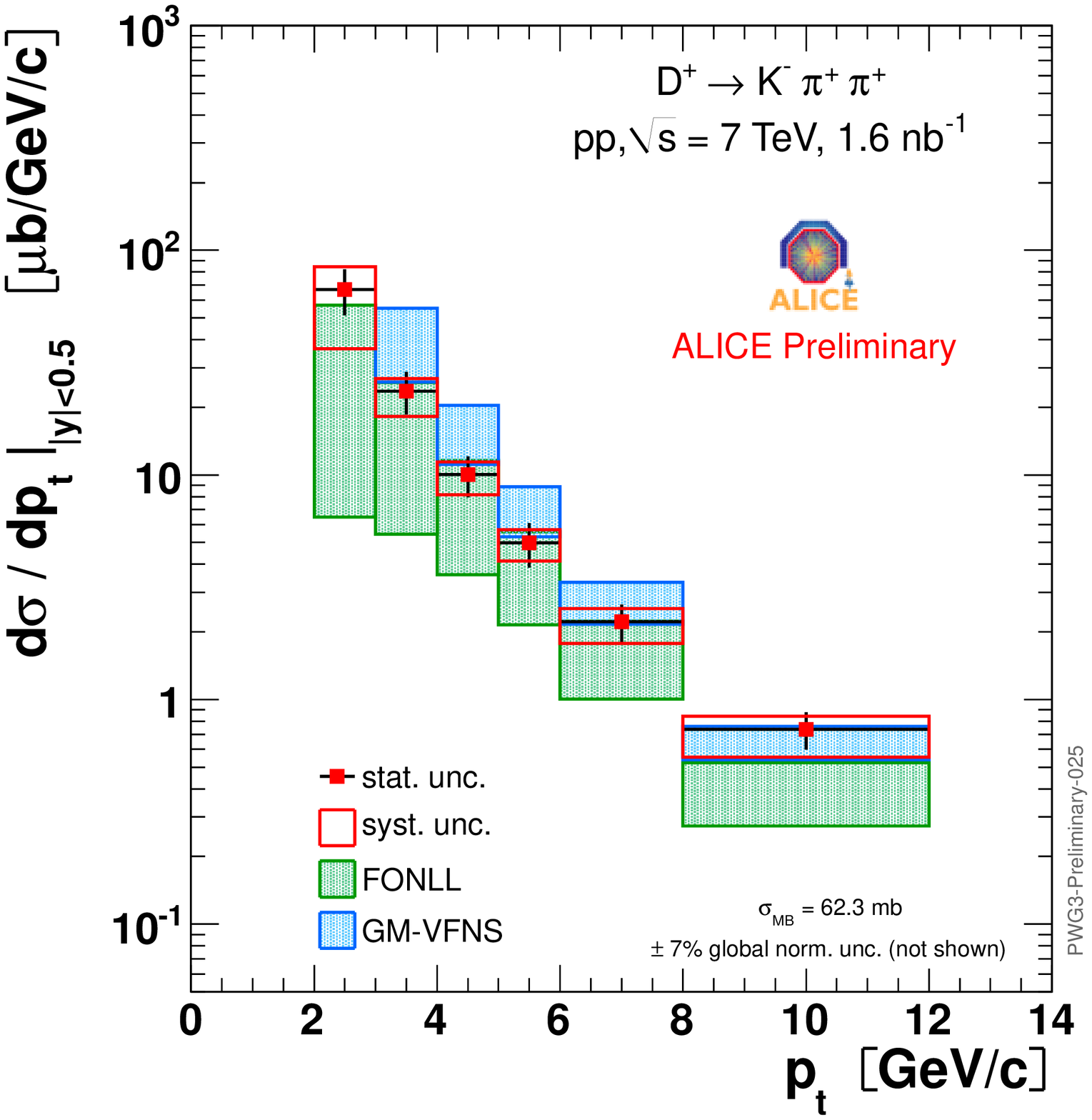}
\includegraphics[width=13pc,height=14pc]{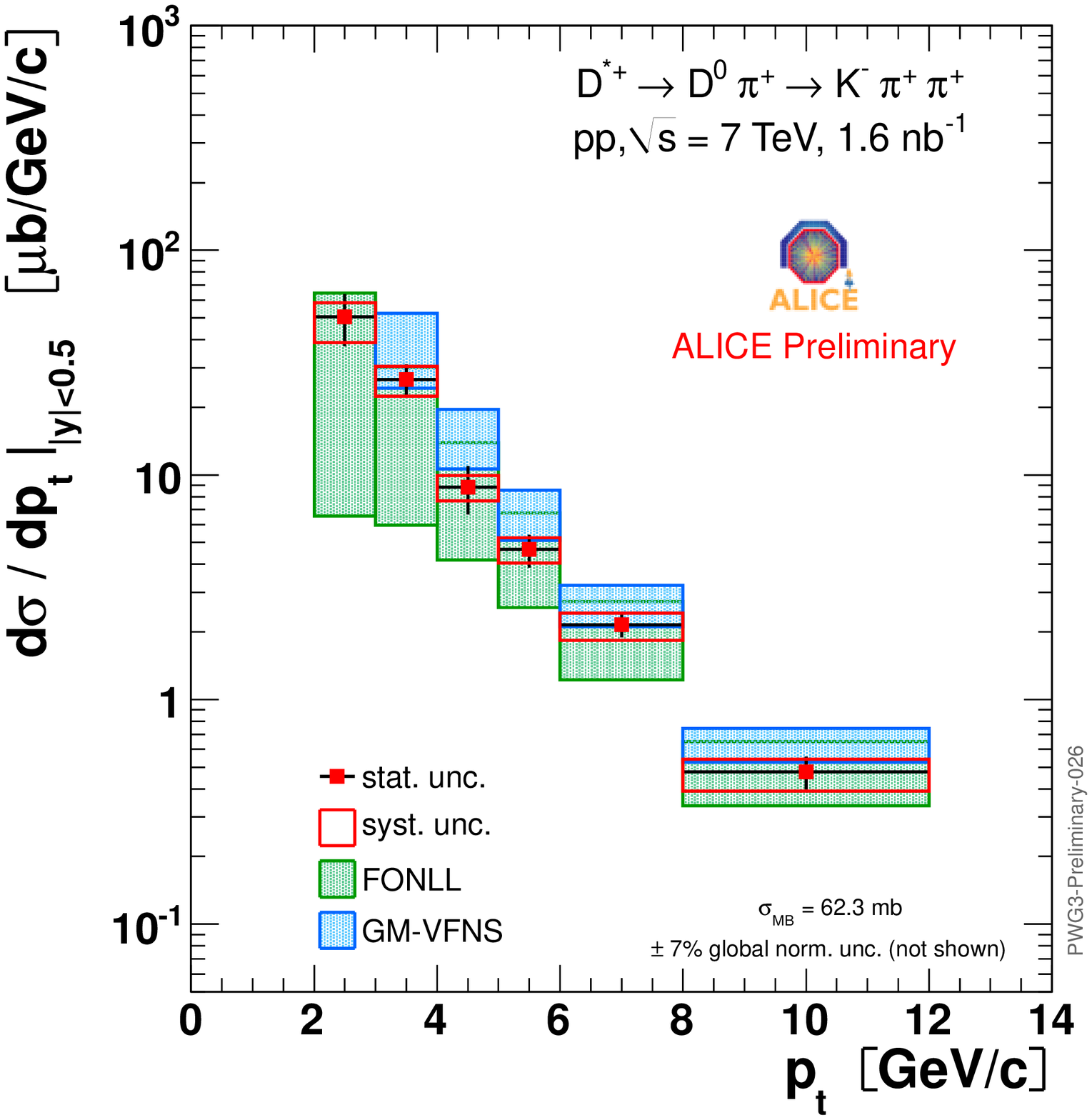}
\caption{ Preliminary $\rm p_{t}$ differential cross section  for $\rm D^{0}$ (left panel) and
$\rm D^{+}$ (middle panel), $\rm D^{*+}$ (right panel) in pp collsions
at $\rm \sqrt{s}$=7
TeV, compared to FONLL \cite{cacciari} and
GM-VFNS \cite{GM} theoretical predictions.}
\label{sigma}
\end{figure}

Fig. \ref{sigma}  shows the  preliminary $\rm  p_t$ differential
cross section for $\rm D^{0}$, $\rm D^{+}$, $\rm D^{*+}$ mesons in pp
collisions at $\rm \sqrt{s}$= 7 TeV,  compared to  two theoretical predictions, FONLL \cite{cacciari} and
 GM-VFNS \cite{GM}. The
 measurements are well-described by both models within their
 theoretical uncertainties.

\begin{figure}[h]
\begin{center}
\includegraphics[width=20pc,height=20pc]{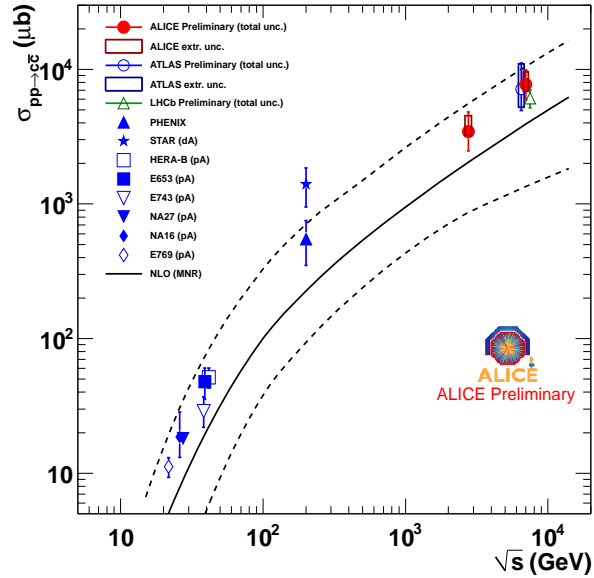}

\caption{Total charm production cross section as a function of
  centre of mass energy for various experiments \cite{Low, ATLAS,
    LHCb, Star data, Phenix}.}
\label{total}
\end{center}
\end{figure}

The same analysis was  repeated for  the pp data sample collected at
$\rm \sqrt{s}$=2.76 TeV, which is the same centre of mass energy as
that of the Pb--Pb. 
 The extrapolation of
the D meson cross section measurements  to the full kinematic
phase space was used to estimate  the total charm production cross
section. This extrapolation was done using FONLL calculations,  
accounting for their uncertainty. The total inclusive charm production
cross section as a function of center of mass energy, shown in
Fig. \ref{total}, illustrates the compatibility of the different
experiments \cite{Low, ATLAS, LHCb, Star data, Phenix}, and  the
reasonable  description of its energy
evolution by the MNR pQCD calculations \cite{pQCD}.

\section{D meson Analysis in Pb--Pb Collisions}

\begin{figure}[h]
\includegraphics[width=20pc,height=20pc]{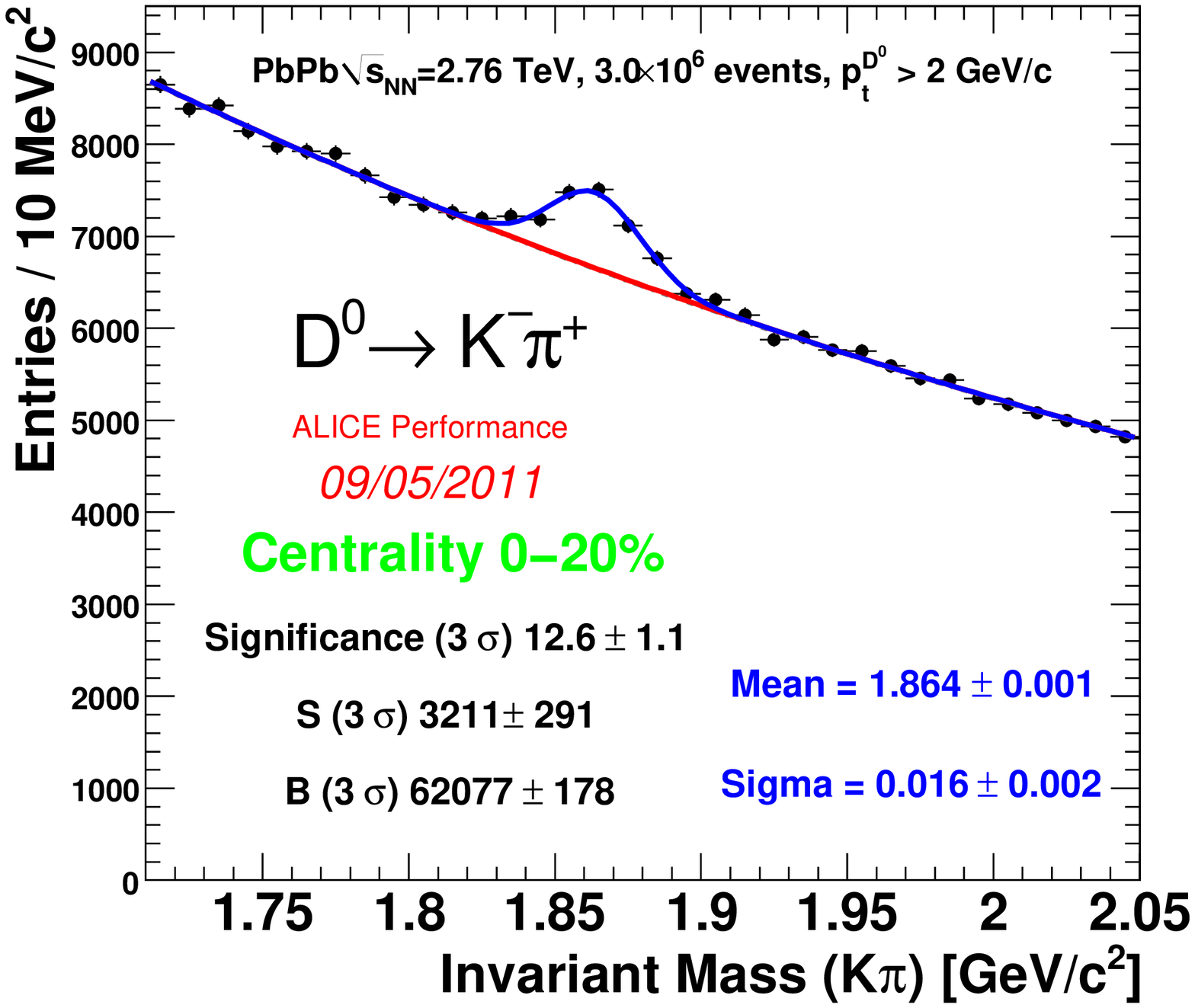}
\includegraphics[width=20pc,height=20pc]{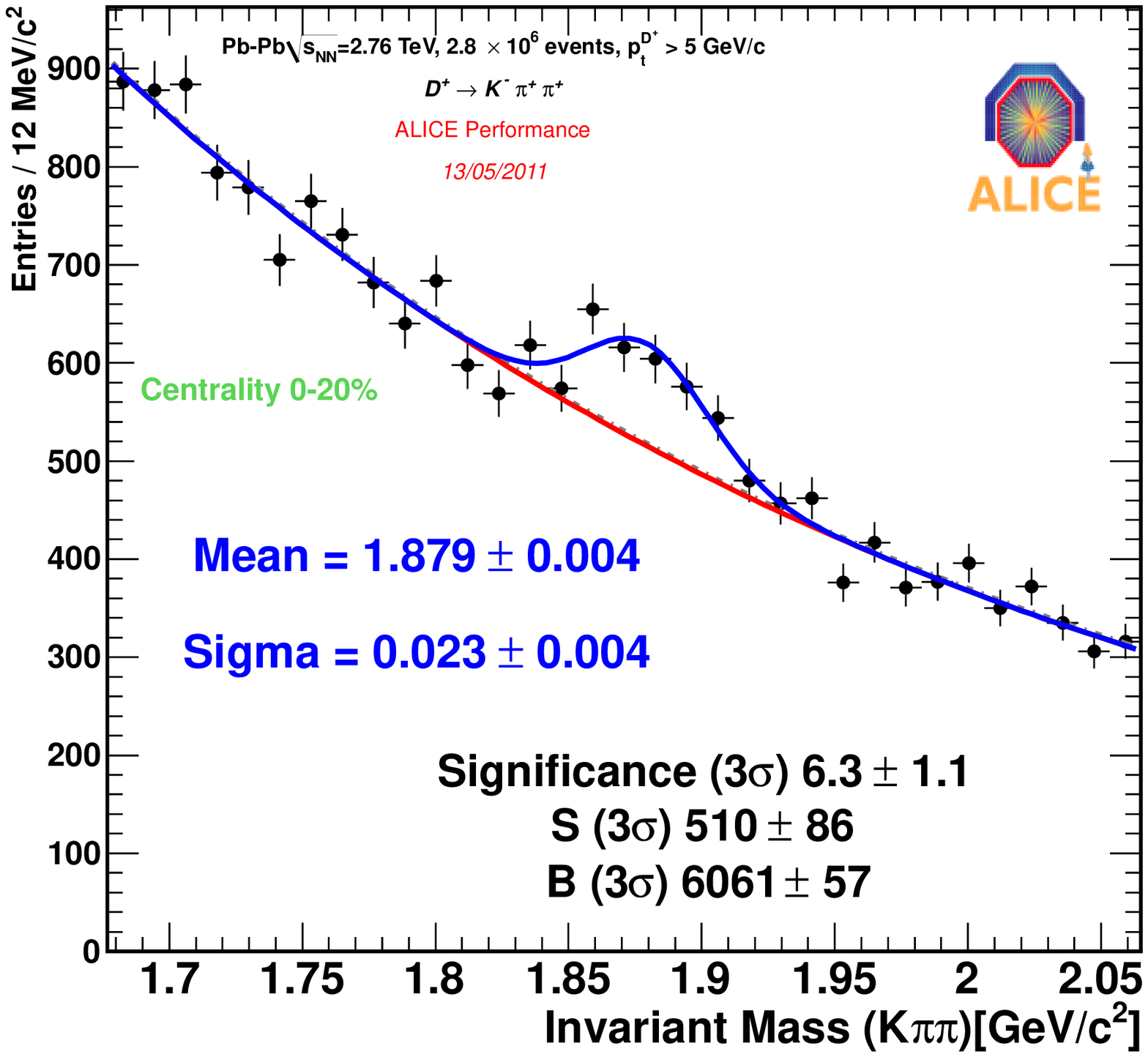}
\caption{Left: Invariant mass distribution from 3 million central Pb--Pb
 collisions for $\rm D^{0}$ candidates with $\rm p_{t} >$ 2 GeV/c,
 Right: $\rm D^{+}$ candidates with $\rm p_{t} >$ 5 GeV/c}
\label{inv}
\end{figure}

A similar analysis strategy has been adopted for Pb--Pb data as for pp,
but with a tighter selection  due to higher combinatorial background.  
The $\rm D^{0} \rightarrow  K^{-} \pi^{+}$  signal was measured in
five $\rm p_{t}$ bins in the range 2 $\rm <p_{t} < $12 GeV/c and the $\rm D^{+} \rightarrow K^{-} \pi^{+}
\pi^{+}$ signal in three bins in the range 5 $\rm < p_{t} <$ 12 GeV/c. The
  invariant mass spectra obtained for $\rm D^{0}$ and $\rm D^{+}$ for the
20$\%$ most central events are shown in Fig. \ref{inv}.

The raw signal is corrected for detector acceptance and efficiency
using Monte Carlo simulations. The contribution of D mesons from
B decays was evaluated using the FONLL prediction. This
contribution is approximately 10-15 $\%$.  The effect of the unknown
nuclear modification of beauty production was included in the systematics. 
\subsection{D meson Nuclear Modification Factor ($\rm R_{AA}$)}

The nuclear modification factor ($\rm R_{AA}$) of  particle
production is defined as: 
\begin{center}

$\rm R_{AA}(p_{t}) =\rm \frac{1}{<N_{coll}>}
. \rm \frac{\rm dN_{AA}/dp_{t}}{\rm dN_{pp}/dp_{t}} =\rm \frac{1}{<T_{AA}>}
. \frac{\rm dN_{AA}/dp_{t}}{\rm d\sigma_{pp}/dp_{t}}$
\end{center}
where  $\rm dN_{AA}/dp_{t}$ is the yield of D-mesons in Pb--Pb and $\rm
d\sigma_{pp}/dp_{t}$ is the cross section measured in pp. $\rm <T_{AA}>$ is the average nuclear overlap function calculated with the Glauber
model in the  centrality range considered.
 
The cross-section  was measured at $\sqrt{s}$ = 7 TeV in the same momentum
range, and must be rescaled to 2.76 TeV to allow a comparison with the Pb--Pb results at the same
energy. This rescaling was done using FONLL  calculation of the  D-meson
$\rm p_{t}$-differential cross
sections at the two energies \cite{scaling}. We assume that  the pQCD scale values (factorization and
renormalization scales) and  the c and b quark masses used in the
calculation do not vary with $\rm \sqrt{s}$. The
theoretical uncertainty in the scaling factor was evaluated by considering the envelope of
the scaling factors resulting from different values of the scales and
heavy quark masses. The uncertainty in the scaling decreases from 25
$\%$ at $\rm p_{t}$ = 2 GeV/c to 10 $\%$ above 10 GeV/c.
The scaling procedure was cross-checked by scaling the
data to $\rm \sqrt{s}$ = 1.96 TeV and comparing with CDF results \cite{CDF}. The
procedure was also validated using the pp run at 2.76 TeV in the
$\rm p_{t}$ range  where the datasets overlap.

\begin{figure}[h]
\includegraphics[width=20pc,height=20pc]{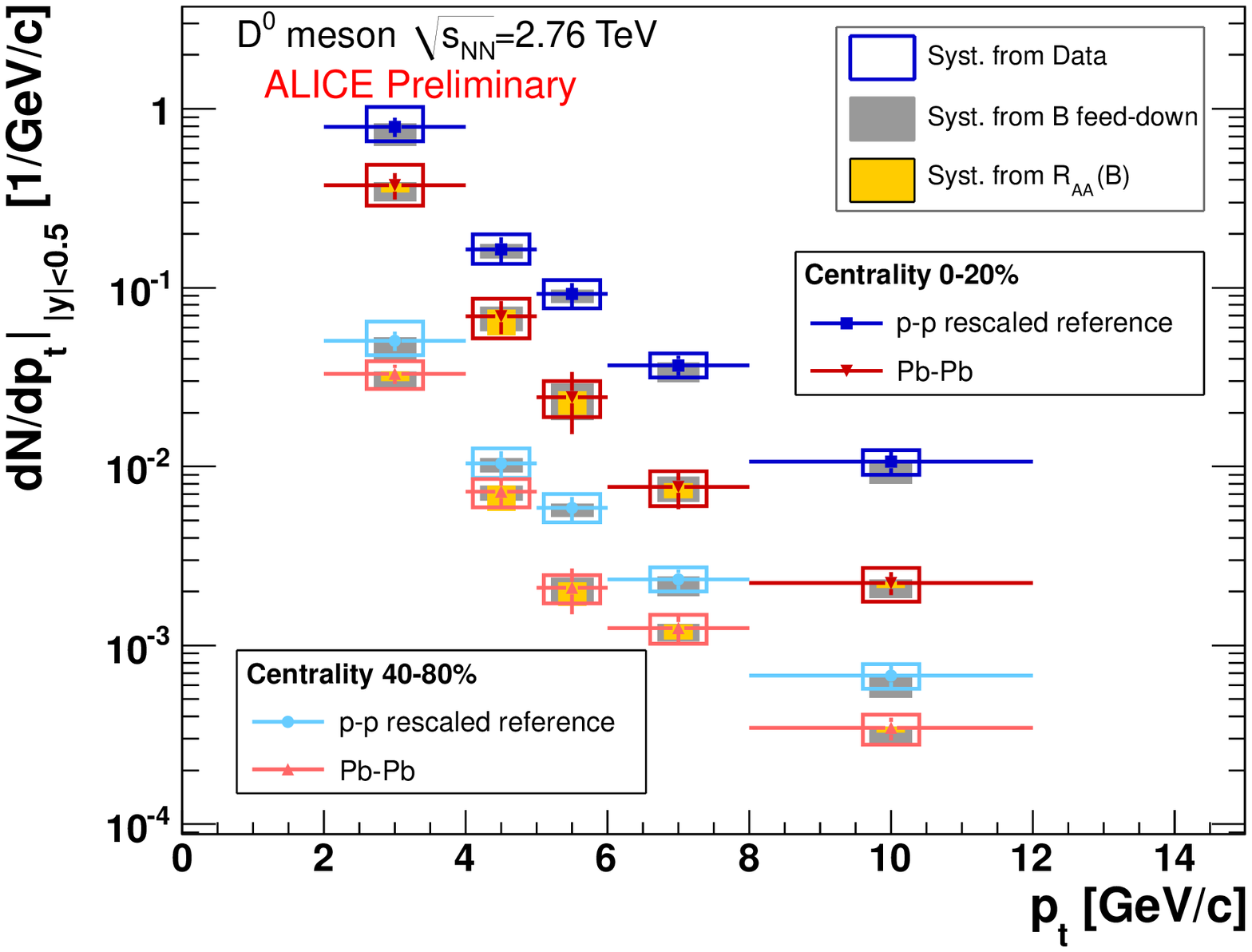}
\includegraphics[width=20pc,height=20pc]{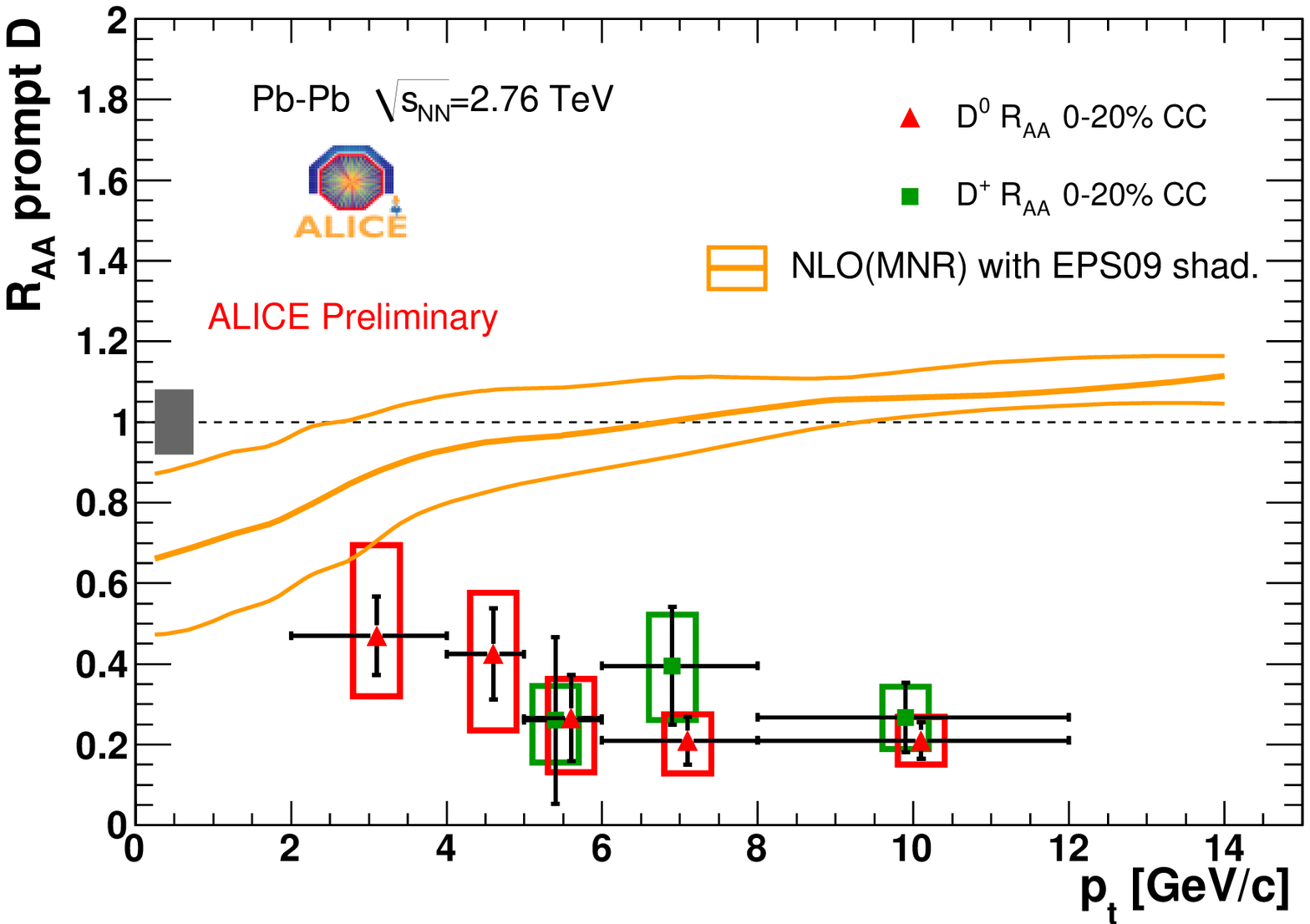}

\caption{Left: $\rm D^{0}$ yield in Pb--Pb central (0-20$\%$)
 and peripheral (40-80$\%$) collisions compared to pp reference
 spectra. Right: $\rm D^{0}$ and $\rm D^{+}$ $\rm R_{AA}$ compared with
$R_{AA}$ of D-meson from pQCD calculation based on MNR code and nuclear PDF from EPS09
parametrization}
\label{pt spectra}
\end{figure}

Fig. \ref{pt spectra}  (left) shows the $\rm D^{0}$  $p_{t}$ spectra
 measured in central (0-20$\%$) and
peripheral (40-80$\%$) Pb--Pb collisions, compared to their respective
reference spectra.

The $\rm D^{0}$  production is suppressed by a factor 4-5 in central events
for $\rm p_{t} >$ 5 GeV/c,
as quantified by the nuclear modification factors shown in Fig.
\ref{pt spectra} (right).   The $\rm D^{0}$ and
$\rm D^{+}$  $\rm  R_{AA}$ agree within errors. Several sources of systematic
uncertainties were considered. They include uncertainties on the
normalization to the total number of events, on the tracking
efficiency,  on  the selection and PID efficiencies, on the signal extraction and on the
$\rm p_{t}$ distribution of D-mesons used to estimate the efficiencies in Monte Carlo
simulations. A systematic uncertainty related to the secondary D-meson
$\rm R_{AA}$ was also taken into account by
varying the value of $R_{AA}$ of D mesons from B decays between 0.3
and 3 times the $\rm R_{AA}$
of prompt D mesons. 

 Nuclear shadowing yields a relatively
small effect for $\rm p_{t} >$5 GeV/c, as shown in Fig. \ref{pt
  spectra} (Right)  by the $\rm R_{AA}$ of D-meson expected
from pQCD calculation based on the MNR code \cite{MNR} and nuclear PDF from EPS09
parametrization \cite{EPS}. Therefore, the observed suppression is  an evidence of in-medium charm quark energy loss.

\begin{figure}[h]
\includegraphics[width=20pc,height=18pc]{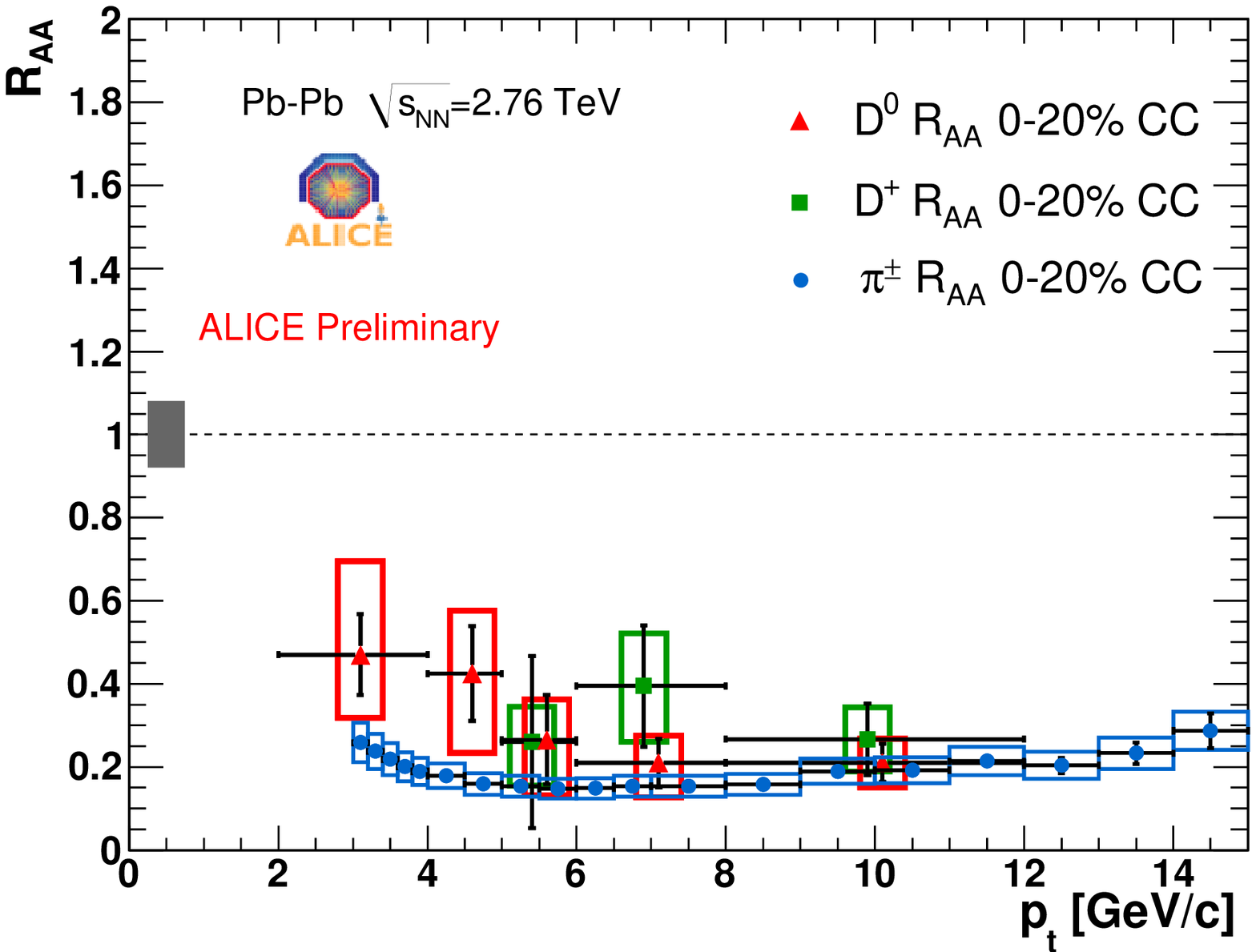}
\includegraphics[width=20pc,height=18pc]{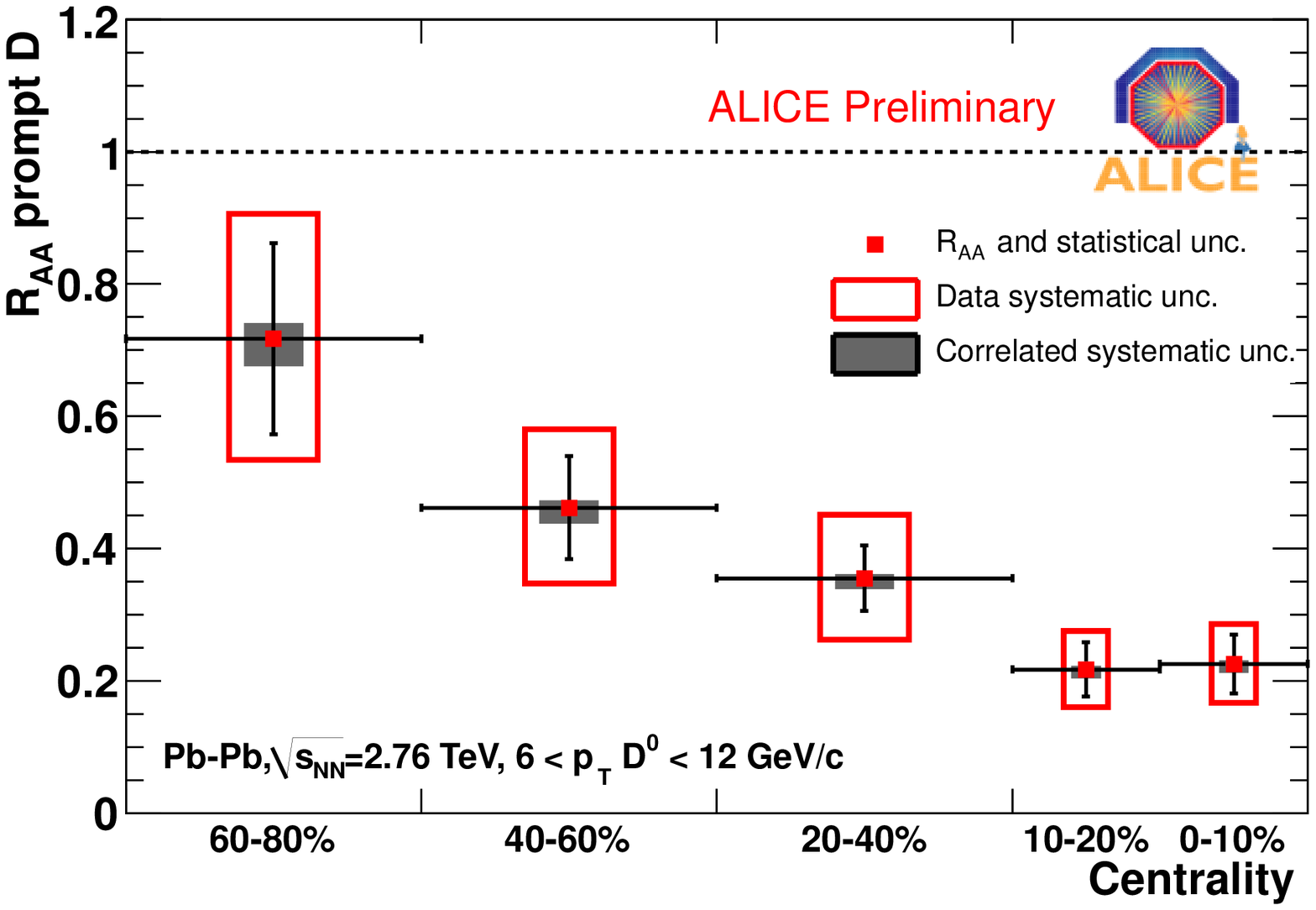}

\caption{Left: Nuclear Modification factor for  $\rm D^{0}$ and $\rm D^{+}$
  compared to that of  charged pions measured in central (0-20$\%$) Pb--Pb collisions. Right: $\rm D^{0}$ $R_{AA}$
as a function of centrality for 6$\rm < p_{t}^{D^0} < $12 GeV/c}
\label{centr dep}
\end{figure}

The D-meson nuclear modification factor was also compared to that of the
charged  pions \cite{pion ref}, shown in Figure \ref{centr dep}
(Left) and the two were found to be  compatible for $\rm p_{t}>$ 5 GeV/c. Below 5 GeV/c,
$\rm R_{AA}$ of pions seems to be smaller than $\rm R_{AA}$ of
D-mesons, but improved statistics are needed to draw firm conclusions.

 The centrality  dependence of  the $\rm D^{0}$  $R_{AA}$ is shown in Fig.
 \ref{centr dep} (right) for 6 $\rm < p_{t} <$ 12 GeV/c.  $R_{AA}$
 decreases from $\sim$ 0.7 in peripheral (60-80$\%$) to $\sim$ 0.2 in central (0-10$\%$) events.

\section{Conclusions}

The present status of the open charm analysis in ALICE has been
shown. For  pp collisions,  we have shown preliminary   measurements of the
$\rm p_{t}$ differential production cross section of the $\rm D^{0}$,  $\rm D^{+}$ and $\rm
D^{*+}$ mesons at central rapidity. The measurements are  well-described by the theoretical
calculations within their uncertainties. In Pb--Pb central collisions,  the
nuclear modification factors of  $\rm D^{0}$ and  $\rm D^{+}$ show a
significant supression at  $\rm p_{t} >$ 5 GeV/c and are  compatible
with the $R_{AA}$ of pions. Below 5 GeV/c, there is a hint of
possible hierarchy in the values of $\rm R_{AA}$ i.e,  $\rm R_{AA}^{D}
> R_{AA}^{\pi} $. The higher statistics, expected from the 2011 Pb--Pb
run should allow this to be confirmed and quantify the difference. In
addition, the comparison data from  p-Pb collisions should
allow  this region to be studied  with greater precision and should disentangle  initial-state nuclear
effects, which could be different for light and heavy 
flavours.


\section*{References}
\begin{thebibliography}{9}
\bibitem{dead-cone} Y. L. Dokshitzer and D.E Kharzeev, Phys. Lett. B
  519,199 (2001).
\bibitem{energy loss} N. Armesto, C. A. Salgado, U. A. Wiedemann, Phys. Rev. D69 (2004) 114003.
\bibitem{CDF} D. Acosta et. al [CDF Coll], Phys. Rev. Lett 91 (2003) 241804.
\bibitem{STAR} A. Adare, et. al [PHENIX Coll], Phys. Rev. Lett 97
  (2006) 252002,  W. Xie [STAR Coll] Proceeding of Hard Probes 2010.
\bibitem{JINST} K. Aamodt et. al, The ALICE experiment at CERN LHC, JINST
 0803:S08002,2008.
\bibitem{Ken} Ken Oyama, for the ALICE Collaboration, proceeding of QM-2011, arxiv:1107.0692v1 [physics.ins-det].

\bibitem{cacciari} M. Cacciari et al., private communication.
\bibitem{GM} B. A. Kniehl et al., private communication.
\bibitem{Low} C. Lourenço, and H.K. Wöhri, Phys. Rept. 433 (2006) 127.
\bibitem{ATLAS} ATLAS collaboration, internal note: ATLAS-CONF-2011-017 (2011).
\bibitem{LHCb} LHCb collaboration, internal note: LHCb-CONF-2010-013 (2010).
\bibitem{Star data} J. Adams et al. (STAR collaboration), Phys. Rev. Lett. 94 (2005) 62301.
\bibitem{Phenix}  A. Adare et al. (PHENIX collaboration),
  arXiv:1005.1627v2 (2010).
\bibitem{pQCD} M. Mangano, P. Nason, and G. Ridolfi, Nucl. Phys. B373 (1992) 295.
\bibitem{scaling} R. Averbeck et al., Reference heavy  flavour cross sections in pp collisions at $\sqrt{s}$=2.76 TeV, using a pQCD-driven
$\sqrt{s}$-scaling of 7 TeV data, arXiv:1107.3243v2 [hep-ph].
\bibitem{MNR} M. Mangano, P. Nason and G. Ridolfi, Nucl. Phys. B373 (1992) 295.
\bibitem{EPS} K. J. Eskola, H. Paukkunen and C. A. Salgado, JHEP 04 (2009) 065.
\bibitem{pion ref} H. Appelshauser, for the ALICE Collaboration,
  proceeding of QM-2011, arxiv:1110.0638v1 [nucl-ex].
\end{thebibliography}
\end{document}